\documentstyle[epsf,preprint,aps]{revtex}

\begin{document}


\preprint{\vbox {\hspace*{\fill} DOE/ER/40762-067\\
          \hspace*{\fill} U. of MD PP \#96-041}} 
\vspace{.5in}

\title{Isospin Violation and Possible Signatures of Disoriented Chiral
Condensates in Ultrarelativistic Heavy Ion Collisions}

\author{Thomas D. Cohen}

\address{Department of 
Physics, University of~Maryland, College~Park, MD~20742-4111}

\maketitle

\begin{abstract}
It is possible that isospin violating dynamical effects are amplified 
due to coherence if a disoriented chiral condensate (or other source
of a coherent pion state) is formed in a heavy ion collison.  It is
shown explicitly 
that  altering the isospin of the coherent  state by order of one unit
can change the number of $\pi^0$'s in the condensate by some finite
fraction of the total number.  A possible signature of such amplified
isospin violating effects is that the ratio of neutral  to total low
$p_T$ pions averaged over many events may differ markedly from 1/3. 
Such a signal would not be washed out if multiple domains of 
coherent pions form.
 \end{abstract}

\newpage
There has been much recent speculation about the possible formation of
regions of coherent or classical pion states in ultrarelativistic
heavy-ion
collisions\cite{dxc1,dxc2,dxc3,dxc4,dxc5,dxc6,dxc7,dxc8,dxc9,dxc10,dxc11
,dxc12,dxc13,dxc14,dxc15,dxc16,dxc17,dxc18,dxc19,dxc20,dxc21,dxc22,dxc23
,dxc24,dxc27}.   
The usual context for such discussions is the
disoriented chiral condensate  (D$\chi$C)
scenario\cite{dxc4,dxc5,dxc6,dxc7,dxc8,dxc9,dxc10,dxc11,dxc12,dxc13,dxc1
4,dxc15,dxc16,dxc17,dxc18,dxc19,dxc20,dxc21,dxc22,dxc23,dxc24,dxc27}.

It is generally assumed that the pions emerging from such states will
be seen at low $p_{T}$ and that the experimental  study of low
$p_{T}$ may provide experimental signatures of such states.
This letter  will focus on  isospin violating effects, which
heretofore have  been neglected in these studies.  Such effects arise
from electromagnetic interactions and due to the mass difference of the
up and down quarks.   In
this letter it will be argued that isospin violating effects
 might radically alter
the phenomenological predictions of a D$\chi$C or other sources of
coherent pion state.   
The coherent nature of
the states may amplify the intrinsically effect of the isospin violation so that
rather modest isospin admixtures into the wave function can have large
consequences.  The effects become particularly interesting  if the
admixtures grow with the number of particles in the condensate.  In
this case, the average ratio of neutral to total, low $p_{\small
T}$, pions  may differ
markedly from 1/3.  This is a signal which will not be washed out
even if multiple uncorrelated regions of coherent pions form.

 A concrete example is the situation in which a single domain
of coherent pions is formed which
has an expectation value of $N$ pions where $N$ is large.    One
natural quantity to study is $R \equiv N_0/N$, where $N_0$ is the number
of $\pi_0$'s in the condensed state and $N$ is the total number of
pions.  If there are no
sources of isospin violation and the condensed state is a pure isoscalar,
then one finds that for large $N$ the probability that the  $R$ will
be found in some interval ${\rm d}R$ is\cite{dxc3,dxc5,dxc6,dxc9}
\begin{equation}
P(R) \, \, {\rm d} R \, =\,  \frac{1}{2 \sqrt{R}} \, \, {\rm d} R \,. \label{pn0}
\end{equation}

Suppose, however, that the isospin carried by the coherent pions is not
exactly zero, how much of an alteration would one find in eq.
(\ref{pn0})?   Since each pion carries a single unit  
of isospin, under normal circumstances one can alter the isospin of
a system of $N$ pions by one unit by simply changing the isospin of a
single pion.   However,
in a coherent state of pions  the isospin of a single pion cannot be
altered in isolation.  The consequence of this is that changing the
isospin of the system by  one unit typically changes the 
number of $\pi_0$ by order of $ a N$ units,
where $a$ is less than 1 and independent of $N$.     For example,  if
the state is isospin zero, the expected number of $\pi_0$'s---$\langle
N_0 \rangle $---is 1/3 $N$,  while, as will be shown below, if the state
is a purely  chargeless  $I=1$ state, then  $\langle N_0 \rangle = 3/5
N$.    Coherence has amplified the effect of changing
a single unit of isospin into  an effect proportional to $N$.
    In this letter, it will be shown that rather small
admixtures of $I \ne 0$ components in the state can lead to large
changes in the distribution of the number
of neutral pions.  

The idea that isospin admixtures of order unity
can yield effects of order $N$ when one has coherent pions
has been considered previously in a different context: such effects are
seen in calculations of $\overline{p} p$ annihilation done in the
context of the  Skyrme model
\cite{nnbar1,nnbar2}.  In that problem the isospin  admixture  is a
consquence of the $\overline{p} p$ initial state .  The
present suggestion is novel in that these effects can come about due to isospin
violating interactions which are normally neglected and, more
importantly, that they can lead to clear signatures of coherent pions
in ultarelativistic heavy ion collisions.
This letter will focus on the problem of  isospin admixtures which keep
the charge fixed at zero.  

Since this letter proposes to study quantum mechanical
admixtures of isospin, some model quantum mechanical state is required.  
The basic underlying assumption is that some particular quantal
mode---{\it i.e.} some particular spatial distribution---has a
macroscopic occupation; the classical limit is simply the large quantum
 number  limit as in the usual correspondence principle.  The
particular choice of a quantum state builds in both the generic
classical (mean-field) features of interest and essentially arbitrary 
fluctations around the mean.  However, so long as one studies the large
 number limit
the effects of the fluctations become small compared to the mean and
the results become independent of the particular choice.
 First consider a classical state
associated with a single scalar (or psuedoscalar) field.  Define
creation and annihilation operators
associated with a spatial distribution of the field given by $\phi(x) $: 
\begin{equation}
a^{\dagger}_\phi \, \equiv \, \int \, {\rm d}^3 x  \, {\rm d}^3 k \, \frac{{\rm
e}^{i k \cdot x} \phi(x) }{2 \omega_k} \, a^{\dagger}_k
\end{equation}
For convenience, it will be assumed the $\phi$ is normalized
( $\int  \, {\rm d}^3 x  \, \phi(x)^2 \, = \, 1$) which implies
$[a_\phi,a_\phi^\dagger]=1$.  
The state associated with $N$ particles in this mode is $| N
\rangle_\phi \, \equiv \, N^{-1/2} \,
(a^\dagger_\phi )^N |0 \rangle$
where $|0\rangle$ is the state of zero occupation.  {\it Any}
supperposition of these states, $\sum_N c_N | N \rangle_\phi$
corresponds to the classical configuaration so long as $ \langle N
\rangle  \gg 1$.  In this letter any such superposition will be refered
to as a coherent state (sometimes refered to as a generalized coherent
state); one common and convenient choice is the standard  coherent
state $e^{-\beta^2/2}e^{-\beta a^\dagger_\phi}|0 \rangle$ \cite{cs}.

For an isotriplet of fields  one with fixed spatial distribution  there
are simply three creation and anhilation operators which satisfy the
commutation relations of   oscillator in the three isospin directions.
 The analogous state for an isotriplet  is  parameterized by a unit
vector $\hat{n}$ which points in some fixed direction in isospace.
The  state with $N$ particles with is given by  
$| \hat{n}, N \rangle_\phi  \equiv  N^{-1/2} (\hat{n} \cdot
\vec{a}^\dagger_\phi)^N |0 \rangle $
 For notational convenience
the subscript $\phi$ indicating the spatial shape will be dropped in
the remainder of this letter.  Again, any supperposition of these
states with different $N$'s gives a coherent state which corresponds to
the classical configuration so long as $ \langle N \rangle \gg 1$.

For any nonzero particle number,
the  coherent state is not an eigenstate of isospin.  To associate a
state of good isospin with the state one must project.
Much of the work on the subject has assumed that the physical state of
interest (if isospin violating dynamics is neglected) is an $I=0$
projection of the coherent state
\cite{dxc6,dxc9}.   For the sake of concreteness and simplicity, this 
picture will be assumed to be correct (neglecting isospin violating
interactions) in this letter. In subsequent work, a more general
case will be considered in which even neglecting isospin violating
interactions   the condensate carries nonzero isospin; however, it is 
strongly correlated with the isospin carried by the other particles in
the system yielding a net isoscalar as in the scenario discussed in
ref. \cite{dxc20}.   

A normalized isospin projected state  of particle  fixed number is
parameterized by   $I$ , $I_3$ and $N$. 
The isospin projected
state is given by \vspace*{-.25in}
 \begin{equation}
| I, I_3, N\beta \rangle\, \equiv \,   {\cal N}_{ I, I_3,N} \int
\,{\rm d}\Omega \, \,  Y_{I, I_3} (\Omega) \, \,| \hat{n}(\Omega), N \rangle
\label{proj}
\end{equation} 
where ${\cal N}_{I,I_3,N}$ is an overall normalization factor.

For an $I=0$  condensate  with $ \langle N \rangle \gg 1$, one obtains
the distribution in eq. (\ref{pn0}) from the states in eq.~(\ref{proj})
with any superpositon of $N$. 
 Suppose instead that the state includes admixtures of chargeless ($I_3=0$)
states with $I \ne 0$:
\begin{equation}
|a, N \rangle = \sum_I a_I |I, I_3=0, N \rangle \label{state}
\end{equation}
 where $\sum_I |a_I|^2 =1$.   Such admixtures are presumed to arise
from  isospin violating dynamics.  The principal result of this letter
is that given such admixtures 
 the probability distribution for $R$, 
the ratio of neutral to total pions in the coherent state
\begin{equation} 
P(R)  = {1\over 2 \sqrt{R}} \, \, 
| \, \sum_I \, a_I \, \sqrt{ 2 I + 1} \, P_I (\sqrt{R})  |^2
\label{pno1}
\end{equation}
where $P_I$ is the $I^{\rm th}$ Legendre polynomial and it is assumed
that $\langle N \rangle \gg1$.  
Equation (\ref{pno1}) is derived from the properties of the  coherent state. 

Equation (\ref{pno1}) implies
small admixtures of  $I \ne 0$ components can alter $P(R)$
signficantly.   The probability
distribution, $P(R)$, is plotted in fig. \ref{p1} for the case  where
the wave function has two components, $I=2$  along with  $I=0$. Even
when the $I=2$ components have a probability of
a few percent, the probalibility
distributions are markedly different from the pure $I=0$ case.  
  
The average value of $R$ in the distribution,
 $\langle R \rangle$ is particularly interesting.   If the state  is a
superposition of
isospin projected states as in eq. (\ref{state}), then
$\langle R \rangle$ is given by: 
\begin{equation}
\langle R \rangle \,=\, \sum_I \left [ \, 2 {\rm Re} (a_I
a^*_{I+2}) \, \frac{( I + 2 ) ( I +1)}{( 2 I +3) \sqrt{( 2 I +1 )
(2I+5)}} \, + \, |a_I|^2 \left (
\frac{1}{3} \, + \, \frac{2}{3} \frac{ I (I+1)}{(2 I -1) (2I+3)} \right) \, \right]
\end{equation}
In fig. \ref{r1} the expectation value of $R$ is shown for states
with $I=0$ and $I=2$ components.  As the
probability of the $I=2$ component increases, $\langle R \rangle$ begins to
differ substantially from 1/3.

The  average value of $R$  is  significant  since it provides a
possible signature of coherence which  survives even if
multiple regions of  coherent pions are formed.   For the pure $I=0$ case, 
$\langle R \rangle = 1/3$ which is exactly the same as one finds in a purely 
statistical model in which pions are emitted without correlations.
The signature which differentiates an $I=0$ source of coherent pions
from  statistical emission are the higher moments of the distribution: a
source of coherent pions will give  larger fluctuations around  the
mean  than purely statistical emission.  Such a
signature will be  diluted if there are  multiple and
uncorrelated regions of coherent pions\cite{dxc17}; multiple regions
 imply a combined distribution which will have  smaller fluctuations
around the mean.  If the number of regions is large the distribution will be
be sharply peaked about $1/3$ and  it will be difficult to
distinguish  from a purely statistical  distribution.

  For $\langle R \rangle \ne 1/3$  to
be a clear signial of isospin violation and coherence, $\langle R \rangle$
must differ  substantially  from 1/3.  There is isospin violation
in incoherent  processes as well.  For example, in a statistical
model, the phase space differences due to the mass differences between
the charge and neutral pions can easily give rise to deviations from
1/3.  One might expect this effect to enhance  $\langle R \rangle$ by
about .01; other sources of isospin violation might contribute
similarly.  Thus, to have a definitive signature one would need to have
$\langle R \rangle$ differ from 1/3 
by several percent.

Whether isospin violations and the formation of regions of coherent
pions alter $\langle R \rangle$ enough to provide a clear signal 
depends on the dynamics of how
the coherent state  gets formed.  Presently, there is no reliable
dynamical model for this.  One can try instead to get insights  from
some very general
considerations.  Suppose the characteristic fractional effect of isospin
violation on some strong interaction process is $\epsilon$, where
$\epsilon$ is perhaps of the order of a couple of percent.  Clearly 
isospin admixtures in a coherent state with an amplitude of $\epsilon$
will   be difficult to distinguish from statistical processes with
isospin violations of the same order.  However, if the
isospin admixtures in the coherent state are characteristically larger
than order $\epsilon$, then $\langle R \rangle$ may differ from 1/3 by
more than order $\epsilon$.  If  the admixtures of $I \ne 0$ components
grow with the number of pions in the condensed state,
then, even for modest $N$, one should expect $\langle R \rangle$
to differ from 1/3 by more than order $\epsilon$.

 It is not implausible that $I \ne 0$ admistures grow with $N$.  First,
note that in  processes which do not involve coherent states the
$I=0$ component typically degrades linearly with $N$ for small $N$---
each additional particle contributes to the isospin breaking. Of
course,  coherence could conceivably suppress this
effect.    However, there is also another suggestive argument.  Consider  a
coherent state projected on fixed particle number $N$ and $I=0$.
Such a state can be written as
\begin{equation}
|N, I=0 \rangle \, = \, \frac{1}{\sqrt{(N+1) !}} \, [  2\, a^\dagger_+
a^\dagger_- \, - \, a^\dagger_0 a^\dagger_0 ]^{N/2} \, | 0 \rangle \,.
\label{aaa}\end{equation}
This state  can be viewed as  a condensed state of $I=0$ pairs of
pions rather than a condensed state of pions; the pair operator given
by $a^\dagger_0 a^\dagger_0\,  -  \, 2\, a^\dagger_+ a^\dagger_-$. 
  It is
natural to suppose that in the
presence of isospin violating interactions the state remains a
condensed state of chargeless pairs of pions, but that pairs are no
longer pure $I=0$:
\begin{equation}
| N, \delta  \rangle \, = \, {\cal N} \,  [  2 \, a^\dagger_+
a^\dagger_- \, - \,  {\rm e}^\delta a^\dagger_0 a^\dagger_0 ]^{N/2} \,|
0 \rangle \,,
\label{delta}
\end{equation} 
where ${\cal N}$ is a normalization constant. 
It  is reasonable to conjecture that  $\delta$ is of order of the
typical isospin violating amplitude and is independent of $N$.     If
this is true,  then  the $I \ne 0$ amplitude  grows with $N$.

If the isospin-mixed-number-projected coherent state is the one
given in eq. (\ref{delta}),  then for $N \gg 1$ the probability
distribution for $R$  is given by
\begin{equation}
P_{\delta N} (R) \, =  \, \frac{ {\rm e}^{N \delta  R} }{2 \sqrt{R}
\int_0^1 \, {\rm d} x \, {\rm e}^ {N \delta x^2}}
\label{pno2}
\end{equation}
The distribution, $P_{\delta N}$ depends only on the product $N \delta$ rather 
than on $\delta$ and $N$ separately:  
isospin violating effects parameterized by $\delta$ clearly grow with $N$.
For $\delta > 0$ the distribution  weighs 
larger values of $R$  heavily; if $N \delta$ is large and positive
the distribution is strongly peaked at $R=1$.  Conversely, negative
$\delta$'s emphasize small values of $R$; if $N \delta$ is
large and negative the distribution is strongly peaked at $R=0$.  

The expectation value of $R$ in  this distribution is  
\begin{equation}
\langle R \rangle_{N \delta} = \frac{\int_0^1 \, {\rm d} x \, x^2 \,
{\rm e}^ {N \delta x^2}}{\int_0^1 \, {\rm d} x \, {\rm e}^ {N \delta x^2}}
\end{equation}
This function is plotted versus $N \delta$ in fig. \ref{r2}.  As
expected the function goes to 1 (zero) for large postive (negative)
values of $N \delta$.  If $\langle R \rangle$ were
nearly 1 or zero would provide a truly spectacular signature.  This 
requires $N |\delta | \gg 1$ which for reasonable
values of $\Delta$ implies $N$ must be several hundred; this is
unlikely in most scenerios.
For relatively small  values of $\delta N$, there is an approximately
linear relationship with $\langle R \rangle_{N \delta} \approx 1/3 + N
\delta/5$.  Thus, if $N$ were  a few tens of particles or larger, one
should have $\langle R \rangle$ sufficiently far from 1/3 to give
a clean signature provided $\delta$ is not unnaturally small.

To summarize: if a source of coherent pions such as  a
disoriented chiral condensate forms in a heavy ion reaction, one
expects isospin violating interactions to alter the distribution of the
ratio of neutral to total pions in the condensate, $R$.   If isospin
violating effects cause  admixtures of $I \ne 0$ components with a
probability of even a few percent, the resultant distribution can be
characteristically different from the $R^{-1/2}/2$ distribution
expected in the absence of isospin violation.  Moreover, if the
probability of $I \ne 0$ components grows with the number of particles
in the condensate as seen, for example, in the  state of eq.
(\ref{delta}), then the expectation value of $R$ can be significantly
different from 1/3 providing a very convincing signature for coherent
pions.   Such a signature will not be washed out if multiple regions of
coherent pions form.   If such an effect were observed it would be
compelling evidence for coherent pions.  

Conversations with A. A. Anselm, S. Nussinov and R. Amado are gratefully
acknowledged.  This work was supported by the U.S. Department of Energy
(grant no. DE-FG02-93ER-40762) and the National Science Foundation
(grant no. PHY-9058487).



\begin{figure}[htb]
\caption {The probability distribution as a function of $R$, the ratio
of charged to neutral pions in the condensed state based on eq.
(\protect{\ref{pno1}}).  The solid curve is based on a pure $I=0$ state, ({\it
i.e.}, $a_j = 0$ for $j \ne 0$).  The dashed and dot-dashed curves
are based on states which are $I=0$ except for an admixture  $I=2$ with
a probability of .03.   The dashed (dot-dashed) curve corresponds to
$a_0$ with the  same  (opposite) sign as  $a_2$.}
\label{p1}
\end{figure}

\begin{figure}[htb]
\caption {\noindent The expectation value of R for coherent states with only $I=0$
and $I=2$ components, as a function of $p_2 = |a_2|^2$, the
probability of the $I=2$ admixture.  The solid (dashed) curve
correspond to $a_0$ with the opposite (same) sign as  $a_2$.}
\label{r1}
\end{figure} 

\begin{figure}[htb]
\caption{The expectation value of R for the states in eq.
(\protect{\ref{delta}}) as a function $N \delta
$.~~~~~~~~~~~~~~~~~~~~~~~~~~~~~~~~~~~~}
\label{r2}
\end{figure}


\begin{thebibliography}{99}

\bibitem{dxc1} A. A. Anselm, JETP  Lett. {\bf 48} (1988) 51.
\bibitem{dxc2} A. A. Anselm, Phys. Lett. {\bf B217} (1989) 169. 
\bibitem{dxc3} A. A. Anselm and  M.G. Ryskin, Phys. Lett. {\bf B266} (1991) 482.
\bibitem{dxc4} J. P.  Blaizot and  A. Krzywicki, Phys. Rev. {\bf D46}
(1992) 246;  {\bf  D50} (1994) 442.
\bibitem{dxc5}  J. D. Bjorken,  SLAC-PUB-5692;  SLAC-PUB-5545 published
in Int. J. Mod. Phys. {\bf A7} (1992) 4189; {Acta. Physica Polonica}
{\bf B23} (1992) 637.
\bibitem{dxc6} K. L. Kowalski and C. C. Taylor,  preprint CWRUTH-92-6 (1992)
\bibitem{dxc7} J. D. Bjorken,  K. L. Kowalski and C. C. Taylor, preprint
SLAC-PUB-6109 (1993).
\bibitem{dxc8} K. Rajagopal and F. Wilczek, {Nucl. Phys.} {\bf B379}
(1993) 395; {\bf B404} (1993) 577.
\bibitem{dxc9} C. Greiner, C. Gong and B. M\"{u}ller, Phys. Lett. {\bf
316} (1993) 226.
\bibitem{dxc10}S. Yu. Khlebnikov,  Mod. Phys. Lett. {\bf A8} (1993) 19901.
\bibitem{dxc11} I. I. Kogan, Phys. Rev. {\bf D48} (1993) 3971; JETP
Lett. {\bf 59} (1994) 307.
\bibitem{dxc12} A. Krzywicki, Phys. Rev. {\bf D48} (1993) 5190.
\bibitem{dxc13} J. P. Blaizot  and  D.  Dyakonov Phys. Lett. {\bf B315} (1993) 226.
\bibitem{dxc14} P. F. Bedaque and A. Das, Mod. Phys. Lett. {\bf A8} (1993) 3151. 
\bibitem{dxc15} A. A. Anselm and Myron Bander, JETP Lett. {\bf 59} (1994) 503.
\bibitem{dxc16} Z.  Huang,  Phys. Rev. {\bf D49} (1994) 16.
\bibitem{dxc17}S. Gavin, A. Gocksch and  R. D. Pisarski,  Phys. Rev.
Lett. {\bf 72} (1994) 2143.
\bibitem{dxc18} Z. Huang and  X. Wang Phys. Rev. {\bf D49} (1994) 4335.
\bibitem{dxc19}S. Gavin and B. M\"{u}ller,  Phys. Lett. {\bf B329} (1994) 486.
\bibitem{dxc20}T. D. Cohen, M. K. Banerjee, M. Nielsen and X. Jin,
Phys. Lett. {\bf B333} (1994) 166. 
\bibitem{dxc21}  Z. Huang, M. Suzuki, X. Wang,  Phys. Rev. {\bf D50} (1994) 2277.
\bibitem{dxc22} R. Amado and I. I.  Kogan, Phys. Rev. {\bf D51} (1995) 190.
\bibitem{dxc23}M. Asakawa, Z. Huang, X. Wang,  Phys. Rev. Lett. {\bf
74} (1995) 3126.
\bibitem{dxc24} M. Martinis, V. Mikuta-Martinis, A. Svarc, J. Crnugelj,
 Fizika {\bf B3} (1994) 197; Phys. Rev. {\bf D51} (1995) 2482.
\bibitem{dxc25}  Z. Huang, M. Suzuki  Phys. Rev. {\bf D52} (1995) 2610;
preprint  LBL-37524 (1995).
\bibitem{dxc26} D. Boyanovsky,  H. J. de Vega,  R. Holman, Phys. Rev.
{\bf D51} (1995)  734. 
\bibitem{dxc27} Zheng Huang,  Phys.Rev. {\bf D49} (1994) 16. 
\bibitem{nnbar1}  R. D. Amado, F.  Cannata, J. P. Dedonder, M. P.
Locher and  B. Shao, Phys. Rev. Lett. {\bf  72} (1994) 72; Phys. Rev.
{\bf C50} (1994) 640. 
\bibitem{nnbar2} B. Shao and R. D. Amado, Phys. Rev. {\bf C50} (1994) 1787.
\bibitem{cs} For a review of coherent states and their applications see
 J. R. Klauder and R. S. Skagerstam, {\it Coherent State, Applications
in Physics and Mathematical Physics} (World Scientific, Singapore 1985).

\end{thebibliography}
\end{document}